\documentclass[aps, amsmath, amssymb, aps, pre, twocolumn, groupedaddress,superscriptaddress,nofootinbib,longbibliography]{revtex4-1}

\usepackage{amsfonts, amsmath, amssymb, xfrac}
\usepackage{enumitem}
\setlist{noitemsep,topsep=0pt,parsep=0pt,partopsep=0pt}
\usepackage{array}
\usepackage[russian,english]{babel}
\usepackage[T2A]{fontenc}
\usepackage[utf8]{inputenc}
\usepackage{graphics}
\usepackage{graphicx}
\usepackage[margin=10pt,parskip=5pt,indention=10pt, lofdepth,lotdepth]{subfig}
\usepackage{listings, xcolor}
\usepackage{bm}
\usepackage{epstopdf}
\usepackage{lineno,hyperref}
\modulolinenumbers[5]

\newenvironment{dfn}[1][Definition]{\begin{trivlist}
\item[\hskip \labelsep {\bfseries #1}]}{\end{trivlist}}

\newtheorem{thm}{Theorem}
\newtheorem{prpsn}[thm]{Proposition}

\newcommand{\mi}{\mathrm{i}}

\begin{document}

\title{Homoclinic orbit and hidden attractor in the Lorenz-like system \\
describing the fluid convection motion in the rotating cavity}

\author{G. A. Leonov}
\affiliation{Faculty of Mathematics and Mechanics, St. Petersburg State University,
Peterhof, St. Petersburg, Russia}
\author{N. V. Kuznetsov}
\email[]{Corresponding author: nkuznetsov239@gmail.com}
\affiliation{Faculty of Mathematics and Mechanics, St. Petersburg State University,
Peterhof, St. Petersburg, Russia}
\affiliation{Department of Mathematical Information Technology,
University of Jyv\"{a}skyl\"{a}, Jyv\"{a}skyl\"{a}, Finland}
\author{T. N. Mokaev}
\affiliation{Faculty of Mathematics and Mechanics, St. Petersburg State University,
Peterhof, St. Petersburg, Russia}
\affiliation{Department of Mathematical Information Technology,
University of Jyv\"{a}skyl\"{a}, Jyv\"{a}skyl\"{a}, Finland}

\date{\today}

\begin{abstract}
In this paper a Lorenz-like system,
describing the process of rotating fluid convection, is considered.
The present work demonstrates numerically that this system, also like
the classical Lorenz system, possesses a
{\it homoclinic} trajectory and a chaotic {\it self-excited} attractor.
However, for considered system, unlike the classical Lorenz one, along with
self-excited attractor a {\it hidden} attractor can be localized.
Analytical-numerical localization of hidden attractor is presented.
\end{abstract}

\maketitle

\section{\label{sec:intro} Introduction}

\begin{figure}[h!]
 \centering
 \includegraphics[width=0.25\textwidth]{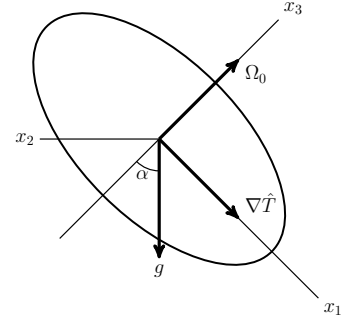}
 \caption{Illustration of the problem setting}
 \label{fig:ellipsoid}
\end{figure}

Consider the following physical problem:
the convection of viscous incompressible fluid flow inside the ellipsoid
\[
 \left(\frac{x_1}{a_1}\right)^2 + \left(\frac{x_2}{a_2}\right)^2 +
 \left(\frac{x_3}{a_3}\right)^2 = 1,
 \quad a_1 > a_2 > a_3 > 0,
\]
under the condition of stationary inhomogeneous external heating.
It is assumed that the ellipsoid together with heat sources rotates
with the constant velocity ${\bf \Omega_0}$ around its axis.
The axis has a constant angle $\alpha$ with the gravity vector $g$.
This vector is stationary with respect to the ellipsoid motion.
The value ${\bf \Omega_0}$ is assumed to be such that the
centrifugal forces can be neglected in
comparison with the influence of gravitational field.
Consider the case when the ellipsoid rotates around the axis $x_3$
and the vector ${\bf g}$ is placed in the plane $x_1 x_3$.
The temperature difference is generated along the axis $a_1$
and a constant $q_0$
is a gradient of this temperature.
(Fig. \ref{fig:ellipsoid}).

Corresponding mathematical model (three-mode model of convection) 
was obtained by Glukhovsky and Dolzhansky
\cite{GlukhovskyD-1980} in the following form
\begin{equation}\label{sys:glukh-golzh}
\begin{cases}
\dot{x} $ = $ A y z + C z - \sigma x, \\
\dot{y} $ = $ -x z + Ra - y, \\
\dot{z} $ = $ x y - z.
\end{cases}
\end{equation}
Here
\begin{align*}
 \sigma &= \frac{\lambda}{\mu}, \qquad Ta = \frac{\Omega_0^2}{\lambda^2}, \quad
 Ra = \frac{g \beta a_3 \hat{q}_1}{2 a_1 a_2 \lambda \mu}, & \\
 A &= \frac{a_1^2 - a_2^2}{a_1^2 + a_2^2} \cos^2 \alpha \, Ta^{-1}, \quad
 C = \frac{2 a_1^2 a_2}{a_3 (a_1^2 + a_2^2)} \sigma \sin \alpha, &\\
 x &= \mu^{-1} \left(\omega_3 + \frac{g \beta a_3 \cos \alpha}{2 a_1 a_2 \Omega_0} q_3 \right), \quad
 y = \frac{g \beta a_3}{2 a_1 a_2 \lambda \mu} q_1, &\\
 z &= \frac{g \beta a_3}{2 a_1 a_2 \lambda \mu} q_2, &
\end{align*}
and $\lambda, \mu, \beta$ are the coefficients of viscosity,
heat conduction, and volume expansion, respectively;
$q_{1}(t)$, $q_{2}(t)$, and $q_{3}(t)$ are the projections of
temperature gradients  on the axes $x_1$, $x_2$ and $x_3$, respectively,
in which case $q_3(t) \equiv 0$;
$\omega_1 (t)$, $\omega_2 (t)$, and $\omega_3 (t)$ are the projections
of the vectors of fluid angular velocities
on the axis $x_1$, $x_2$, and $x_3$, in which case
\[
 \omega_1 = - \frac{g \beta a_3}{2 a_1 a_2 \Omega_0} \cos \alpha q_1, \quad
 \omega_2 = - \frac{g \beta a_3}{2 a_1 a_2 \Omega_0} \cos \alpha q_2.
\]
The parameters $\sigma$, $Ta$, and $Ra$ are the
Prandtl, Taylor, and Rayleigh numbers, respectively.

After two sequential transformations
\begin{align*}
 & x \to x, \quad y \to C^{-1}y, \quad z \to C^{-1}z, \\
 & x \to x, \quad y \to R - \frac{\sigma}{a_0 R+1} z, \quad z \to \frac{\sigma}{a_0 R+1} y,
\end{align*}
one obtains the following system
\begin{equation}
\begin{cases}
 \dot{x} $ = $ - \sigma(x-y) - ayz\\
 \dot{y} $ = $ rx-y-xz \\
 \dot{z} $ = $ -z+xy,
\end{cases}\label{sys:lorenz-general}
\end{equation}
where $a_0 = A / C^2$, $R = Ra \, C$,
\begin{equation}
 a = \frac{a_0 \sigma^2}{(a_0 R +1)^2}, \quad r = \frac{R}{\sigma}(a_0 R + 1).
 \label{sys:lorenz-general:param}
\end{equation}

In the case $a=0$ system \eqref{sys:lorenz-general} coincides
with the classical Lorenz system \cite{Lorenz-1963}.
For the first time,  system \eqref{sys:lorenz-general} with
parameters $r, \, \sigma > 0$ was considered in \cite[1978]{Rabinovich-1978}.
After the linear change of variables \cite{LeonovB-1992}
this system can be reduced to the Rabinovich system, describing
the waves interaction in plasma
\cite{PikovskiRT-1978, XieZh-2003, Zhang-2003, ZhangMWWY-2014}.
As is shown in \cite{LeonovB-1992} system \eqref{sys:lorenz-general}
describes the following physical processes:
the flow of fluid convection inside the rotating ellipsoid
\cite{GlukhovskyD-1980},
the rotation of rigid body in viscous fluid \cite{Denisov-1989},
the gyrostat dynamics \cite{Glukhovsky-1982,Glukhovsky-1986},
the convection of a horizontal layer of fluid making the harmonic oscillations \cite{ZaksLCh-1983},
and the model of Kolmogorov's flow \cite{DovzhenkoD-1987}.
In \cite{EvtimovPS-2000} for system \eqref{sys:lorenz-general}
in the case $\sigma = \pm ar$
a detailed analysis of the equilibria stability and asymptotic behavior
of trajectories is given and the values of parameters are obtained
for which system \eqref{sys:lorenz-general} is integrable.
Remark also the works \cite{PanchevSV-2007,LiaoTA-2010},
in which the analytical and numerical study of
some generalizations of system \eqref{sys:lorenz-general}
and similar systems is presented.
In addition, in \cite{AkhtanovZZ-2013} system
\eqref{sys:lorenz-general} was used to describe a specific
scenario of transition to chaos in low-dimensional dynamical systems ---
gluing bifurcations.

Note that the Glukhovsky-Dolzhansky system is sufficiently different  from the
classical Lorenz system. In the Lorenz system, the flow of the two-dimensional
convection is considered only. In the Glukhovsky-Dolzhansky system,
the flow of the three-dimensional convection
is considered which can be interpreted as one of the models of ocean flows
\cite{GlukhovskyD-1980}.

In what follows system \eqref{sys:lorenz-general} will be considered
under the condition that the parameter $a$ is positive.
In this case if $r < 1$, then \eqref{sys:lorenz-general}
has a unique equilibrium ${\bf \rm S_0} = (0,0,0)$, which is
globally asymptotically Lyapunov stable
\cite{LeonovB-1992,BoichenkoLR-2005}.
If $r > 1$, then \eqref{sys:lorenz-general}
posesses three equilibria: ${\bf \rm S_0} = (0,0,0)$ and
\begin{equation}
 {\bf \rm S_{1,2}} = (\pm x_1, \, \pm y_1, \, z_1). \label{eq:equil_s12}
\end{equation}
Here
\[
 x_1 = \frac{\sigma \sqrt{\xi}}{\sigma + a \xi}, \quad
 y_1 = \sqrt{\xi}, \quad
 z_1 = \frac{\sigma \xi}{\sigma + a \xi},
\]
and the number $\xi$ is defined as
\[
 \xi = \frac{\sigma}{2 a^2} \left[ a (r-2) - \sigma + \sqrt{(\sigma - ar)^2 + 4a\sigma} \right].
\]
The stability of equilibria $S_{1,2}$ depends on
the parameters  $\sigma$, $r$, $a$ (see. Sec. \ref{sec:stab-anal}).

For system \eqref{sys:lorenz-general} with the fixed $\sigma$,
 $a$ (or with the fixed $\sigma$ only)
it is possible to observe a classical scenario of transition to the chaos
similar to scenario in the Lorenz system \cite{Sparrow-1982}.
To demonstrate this, for system \eqref{sys:lorenz-general}
with the fixed parameters  $\sigma$ and $a$ and
increasing parameter $r > 1$, a homoclinic trajectory and a self-excited
chaotic attractor are obtained numerically.
Unlike the Lorenz system, for system \eqref{sys:lorenz-general}
it is also possible to localize a hidden chaotic attractor.

\section{\label{sec:homoclin} Homoclinic orbit in the Lorenz-like system}

Denote by ${\bf \rm x}(t)$ a trajectory of system \eqref{sys:lorenz-general}
starting at a certain initial point.
In order to compute numerically a homoclinic trajectory
(\( \lim_{t \to +\infty} {\bf \rm x}(t) = \lim_{t \to -\infty} {\bf \rm x}(t) = S_0 \)),
one integrates system \eqref{sys:lorenz-general}
with the initial data ${\bf \rm x}_0$
from a $\delta$-vicinity of the saddle point $S_0$ and
its one-dimensional unstable manifold $W^u (S_0)$
that corresponds to a positive eigenvalue of the Jacobian matrix $J$
at the saddle point $S_0$.
For some parameters of system \eqref{sys:lorenz-general} this trajectory
after a certain time intersects a two-dimensional plane $M$
spanned on the eigenvectors that correspond to negative eigenvalues of $J$.
The parameters are chosen in such a way that the point of intersection
belongs to $\delta$-vicinity of $S_0$.
\begin{figure}[!hb]
 \centering
 \includegraphics[width=0.4\textwidth]{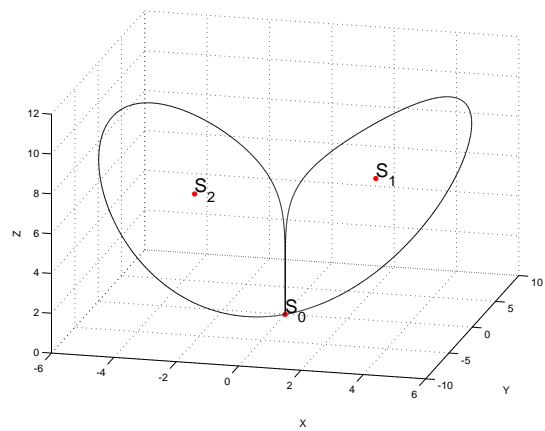}
 \caption{\label{fig:homoclinics} Approximation of the homoclinic butterfly
for system
 \eqref{sys:lorenz-general}.}
\end{figure}

\begin{figure*}[!ht]
 \centering
 \captionsetup{justification=centering}
 \subfloat[$r = 7.44$
 ]{
 \label{fig:homoclin:before-biff}
 \includegraphics[width=0.33\textwidth]{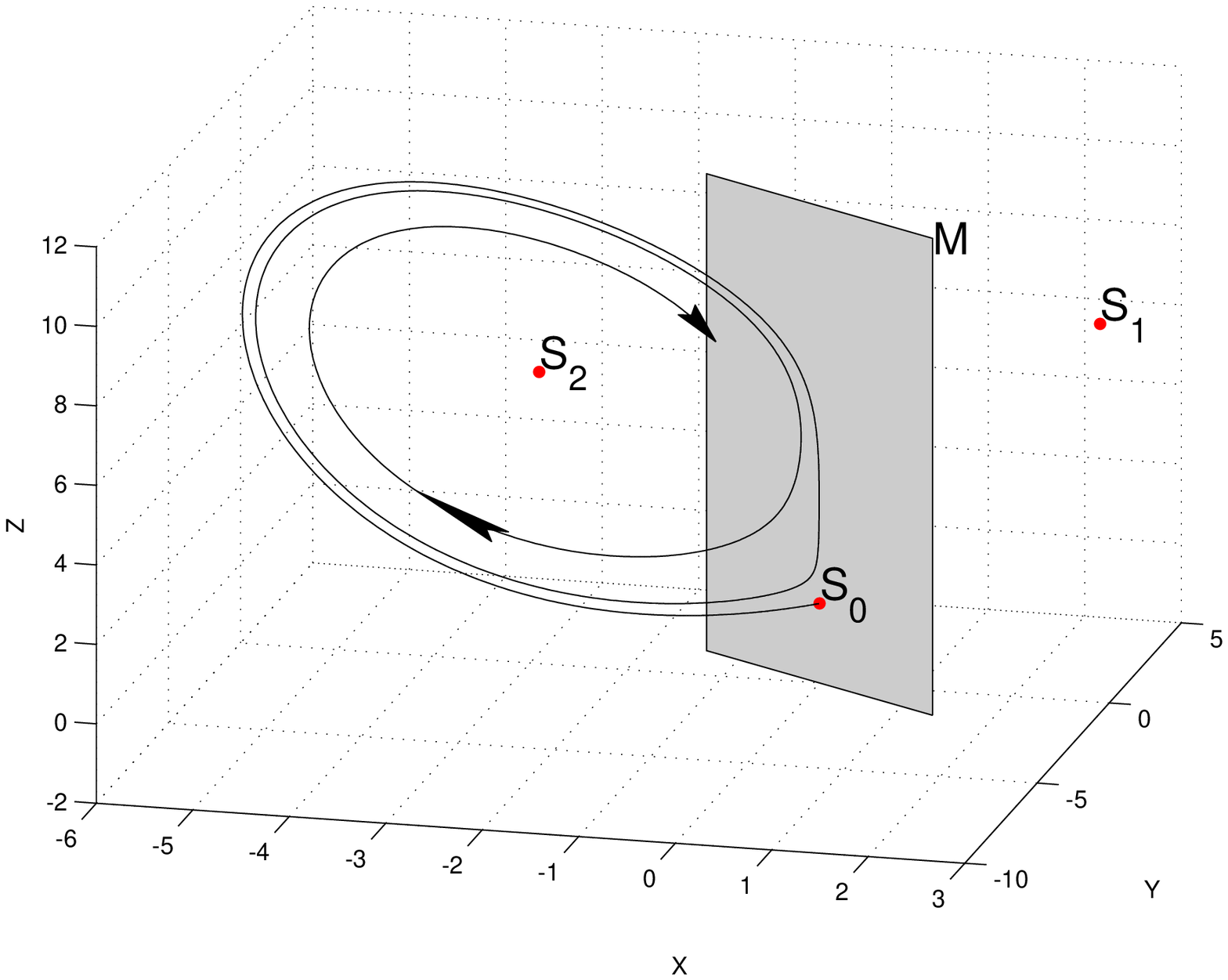}
 }
 \subfloat[$r = 7.4430045820796753...$
 ]{
 \label{fig:homoclin:biff}
 \includegraphics[width=0.33\textwidth]{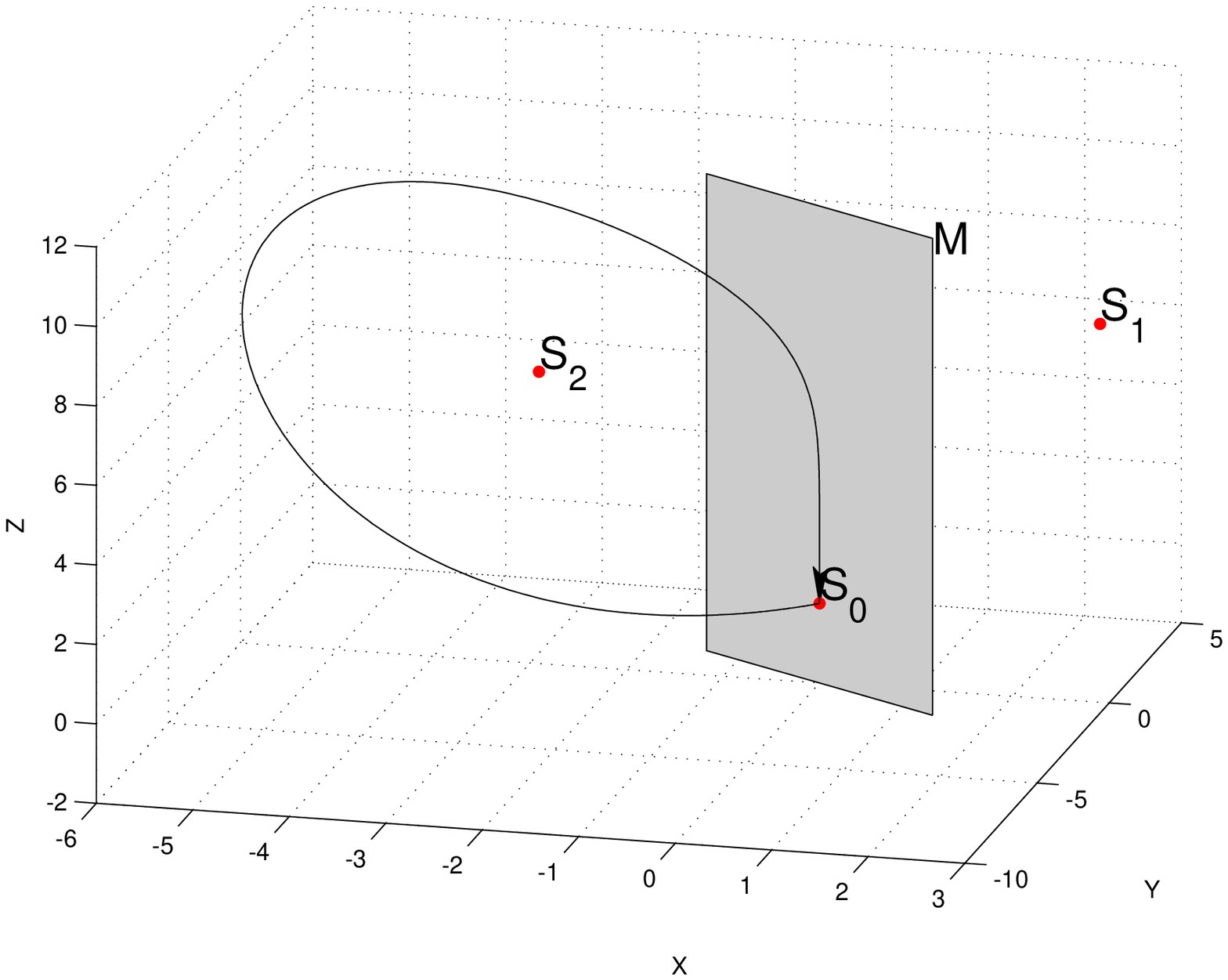}
 }
 \subfloat[$r = 7.445$
 ]{
 \label{fig:homoclin:after-biff}
 \includegraphics[width=0.33\textwidth]{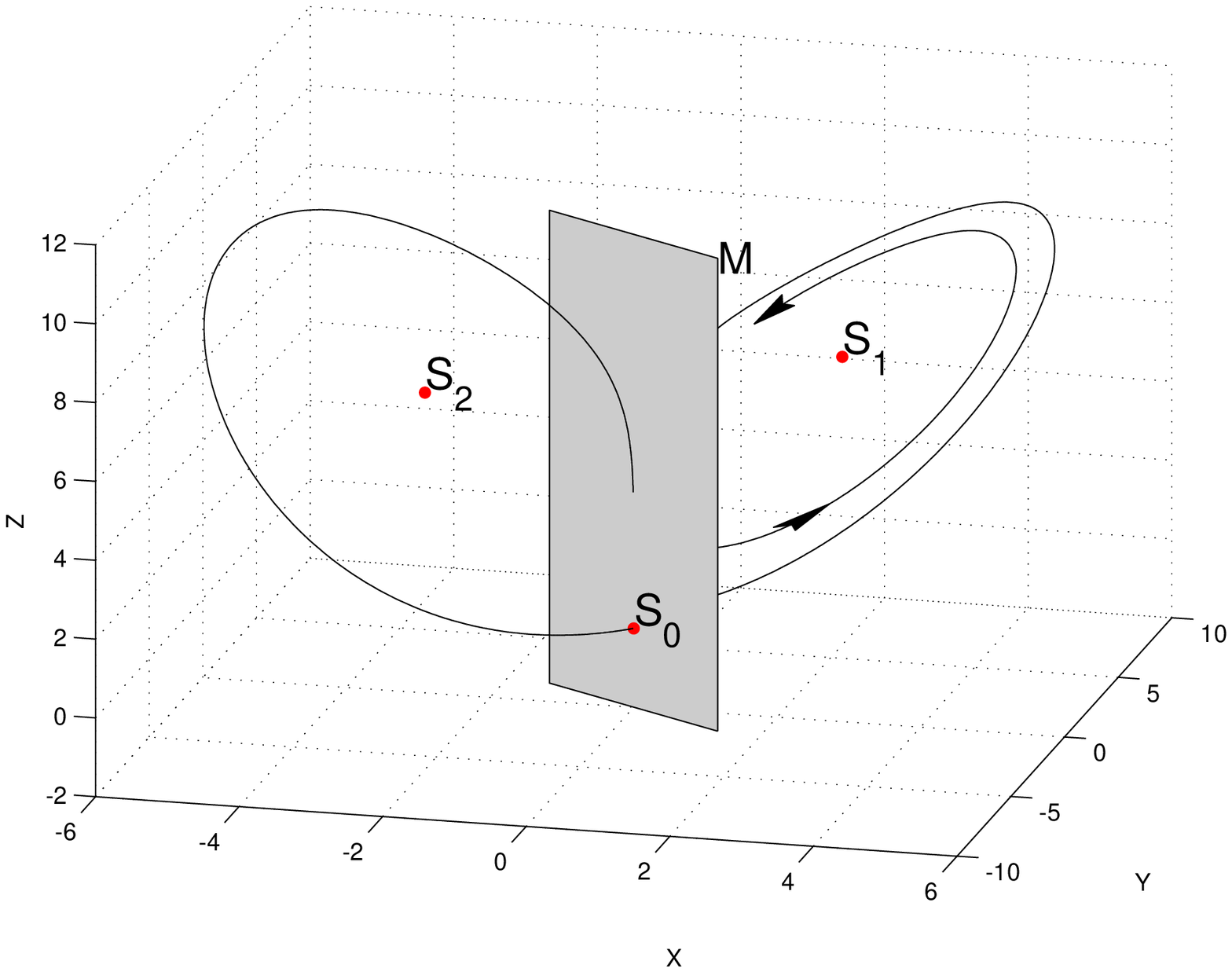}
 }
 \caption{\label{fig:homoclin} The birth of a homoclinic orbit
in system  \eqref{sys:lorenz-general} with $\sigma = 4$, $a = 0.0052$,
$r \in [7.44, 7.445]$, $\delta = 0.01$.}
\end{figure*}

Let us fix the parameters: $\sigma = 4$ and $a = 0.0052$
(such values were considered in \cite{GlukhovskyD-1980}).
For $r = 7.44$ there is no intersection of the trajectory ${\bf \rm x}(t)$
with the plane $M$ (see Fig. \ref{fig:homoclin:before-biff}) and
for $r = 7.445$ the intersection occurs (see Fig. \ref{fig:homoclin:after-biff}).
So, there exists an intermediate value $r^{*} \in [7.44, 7.445]$
for which one can get the approximation of homoclinic orbit $r^{*} = 7.4430045820796753...$
(see Fig. \ref{fig:homoclin:biff}).
Note that the approximation for the symmetric homoclinic
orbit can be obtained by the choose in the computational procedure the
symmetric (with respect to $S_0$) initial data (see Fig. \ref{fig:homoclinics}).
From an analytical point of view, the existence of homoclinic trajectory can be justified by
{\it Fishing principle} \cite{Leonov-2012-PLA,Leonov-2013-IJBC,Leonov-2014-ND}.
The Fishing principle is based on the construction of a special two-dimensional manifold
such that the separatrix of the saddle point intersects or
does not intersect the manifold for two different values of a system parameter.
The continuity implies the existence of some intermediate
value of the parameter for which the
separatrix touches the manifold.
According to construction the only possibility for separatrix
is to touch the saddle and thus, one can numerically localize
the birth of the homolcinic orbit by changing the variable parameter.

\section{\label{sec:attractor} Chaotic attractor in the Lorenz-like system}

\subsection{\label{sec:stab-anal} Local stability analysis and computation of attractors}

Let us study the stability of equilibria $S_1$, $S_2$ of system \eqref{sys:lorenz-general}.
By the Routh-Hurwitz criterion, one can get the following
\begin{prpsn}\label{prpn:gen_lorenz:stability}.
If $\sigma > 2$ and the parameters  $r$ and $a$ satisfy the inequality
\begin{equation}
 p_3(\sigma, A) \, r^3 + p_2(\sigma, A) \, r^2 + p_1(\sigma, A) \, r + p_0(\sigma, A) < 0,
 \label{eq:ineq:ra1}
\end{equation}
where
\begin{align*}
 p_3(\sigma, A) &= a^2 \sigma^2 (\sigma -2), \\
 p_2(\sigma, A) &= - a \left(2\sigma^4 - 4\sigma^3 - 3 a \sigma^2 + 4 a \sigma + 4 a\right), \\
 p_1(\sigma, A) &= \sigma^2 \left(\sigma^3 + 2(3a - 1)\sigma^2 - 8 a \sigma + 8 a\right), \\
 p_0(\sigma, A) &= - \sigma^3 \left(\sigma^3 + 4\sigma^2 - 16a\right),
\end{align*}
then the equilibria $S_{1,2}$ are stable.
\end{prpsn}

Let us choose the parameter $\sigma = 4$
and, as in \cite{GlukhovskyD-1980},
construct the domains of stability of the equilibria $S_{1,2}$
of system \eqref{sys:lorenz-general} in dependence on
the values of parameters  $a$ and $r$.
Then inequality \eqref{eq:ineq:ra1} takes the form
\begin{equation}
 8 a^2 r^3 + a(7 a - 64) r^2 + (288 a + 128) r + 256 a - 2048 < 0.
 \label{eq:ineq:ra2}
\end{equation}
\begin{figure}[ht!]
 \centering
 \includegraphics[width=0.35\textwidth]{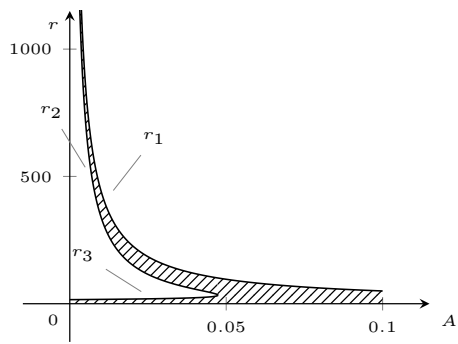}
 \caption{Domain of stability of the equilibria $S_{1,2}$ of system \eqref{sys:lorenz-general}
 for $\sigma = 4$.}
 \label{fig:gen_lorenz:regimes}
\end{figure}
For $0 < a < a^*$, where $a^* = 0.04735...$,
there are three real roots: $r_1 (a) > r_2 (a) > r_3 (a)$,
for $a = a^*$ two real roots: $r_1 (a)$ and $r_2(a) = r_3(a)$, and
for $a > a^*$ one real root: $r_1 (a)$.

Thus, for $0 < a < a^*$ the equilibria $S_{1,2}$
are stable for $r < r_3 (a)$ and $r_2 (a) < r < r_1 (a)$
and they are unstable in the converse case;
for $a > a^*$ the equilibria $S_{1,2}$ are stable for $r < r_1 (a)$
(see, Fig. \ref{fig:gen_lorenz:regimes}).


An oscillation in a dynamical system can be easily localized numerically
if the initial conditions from its open neighborhood
lead to the long-time behavior
that approaches the oscillation.
Thus, from a computational point of view it is natural to suggest the following classification of attractors,
based on the simplicity of finding the basin of attraction in the phase space:

\begin{dfn}\cite{KuznetsovLV-2010-IFAC,LeonovKV-2011-PLA,LeonovKV-2012-PhysD,LeonovK-2013-IJBC}
 An attractor is called a \emph{hidden attractor} if its
 basin of attraction does not intersect with
 small neighborhoods of equilibria,
 otherwise it is called a \emph{self-excited attractor}.
\end{dfn}

\subsection{\label{sec:attractor:gen-lorenz:self-ex}
Self-excited attractor in the Lorenz-like system}

For a \emph{self-excited attractor} its basin of attraction
is connected with an unstable equilibrium
and, therefore, self-excited attractors
can be localized numerically by the
\emph{standard computational procedure},
in which after a transient process a trajectory,
started from a point of an unstable manifold in a neighborhood
of an unstable equilibrium,
is attracted to the state of oscillation and traces it.
Thus self-excited attractors can be easily visualized.

Using the constructed domain of stability (\ref{fig:gen_lorenz:regimes}),
 one considers a qualitative behavior of trajectories
 of system \eqref{sys:lorenz-general} for the fixed $\sigma = 4$,
$a = 0.0052$, and $r \in (16.4961242,\, 690.6735024)$.
For system \eqref{sys:lorenz-general}
the parameter $r = 687.5$ is chosen.

For the above parameters the eigenvalues of equilibria of system \eqref{sys:lorenz-general}
are the following
\begin{align*}
 S_0 : \quad & 49.9619, \quad -1, \quad -54.9619 \\
 S_{1,2} : \quad & 0.0295 \pm 10.729 \, \mi, \quad -6.0591
\end{align*}
\begin{figure}[ht!]
 \centering
 \subfloat[
 {
 \scriptsize Attractor for initial conditions in the neighborhood of $S_0$
 }
 ] {
 \label{fig:gen_lorenz:attr:se1}
 \includegraphics[width=0.4\textwidth]{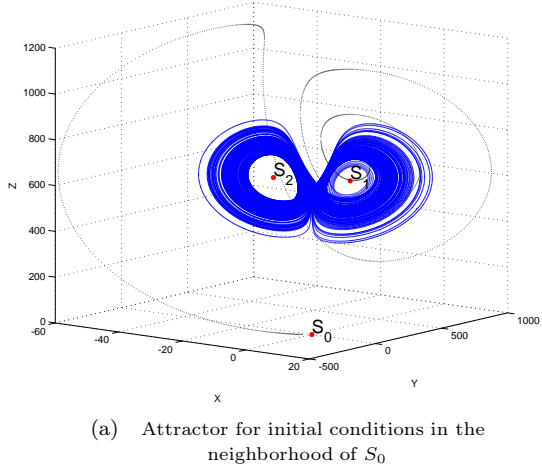}
 }

 \subfloat[
 {
 \scriptsize Attractor for initial conditions in the neighborhood of $S_1$
 }
 ] {
 \label{fig:gen_lorenz:attr:se2}
 \includegraphics[width=0.4\textwidth]{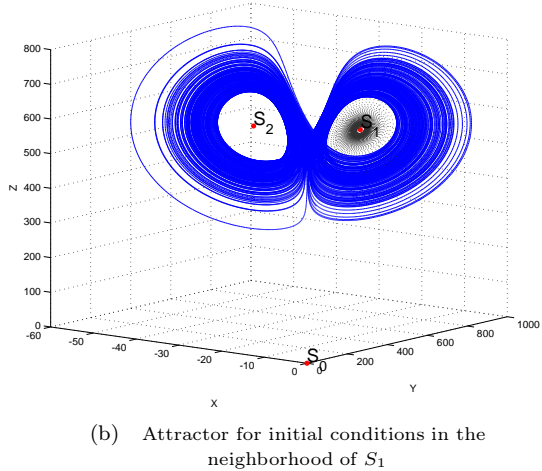}
 }
 \caption{
 A self-excited attractor of system \eqref{sys:lorenz-general} for
 $r = 687.5$, $\sigma = 4$, $a = 0.0052$.
 }
 \label{fig:gen_lorenz:attr:se}
\end{figure}
Thus, the equilibrium $S_0$ is a saddle
and $S_{1,2}$ are saddle-focuses.
Having taken an initial point on the unstable manifold
of one of equilibria $S_{0,1,2}$ (Fig. \ref{fig:gen_lorenz:attr:se}),
one can easily be vizualized a self-excited chaotic attractor
by standard computational procedure.

For $r \in (690.6735024, \, 830.4169122)$ the equilibria $S_{1,2}$
become stable and the trajectories, starting from the neighborhood
of  equilibrium $S_0$, are attracted to $S_1$ or $S_2$.
The question arises whether there exists a hidden chaotic attractor
in system \eqref{sys:lorenz-general} for such values of parameters?
Next for the computation of hidden attractor in system \eqref{sys:lorenz-general}
a special numerical procedure is considered.

\subsection{\label{sec:attractor:gen-lorenz:hidden} Hidden attractor
in the Lorenz-like system}

For a hidden attractor its basin of attractor is not related with
unstable equilibria.
The hidden attractors, for example, are the attractors in the systems
with no equilibria or with only one
stable equilibrium (a special case of multistable
systems and coexistence of attractors).
Recent examples of hidden attractors can be found in
\citep{LeonovKKSZ-2014,ZhusubaliyevM-2014,PhamJVWG-2014,PhamRFF-2014,WeiWL-2014,LiSprott-2014-IJBC,WeiML-2014-TJM-cu,
KuznetsovKMS-2014,LiZY-2014-cu,ZhaoLD-2014,LaoSJS-2014-cu,ChaudhuriP-2014-cu}.
Multistability is often an undesired situation in many applications
but the coexisting self-excited
attractors can be found by the standard computational procedure.
In contrast, there is no regular way to predict the existence
or coexistence of hidden attractors in system.
Note that one cannot guarantee the localization of attractor
by the integration of trajectories with random
initial data (especially for multidimensional systems)
since its basin of attractor may be very small.
\begin{figure}[!ht]
	\centering
	\includegraphics[width=0.4\textwidth]{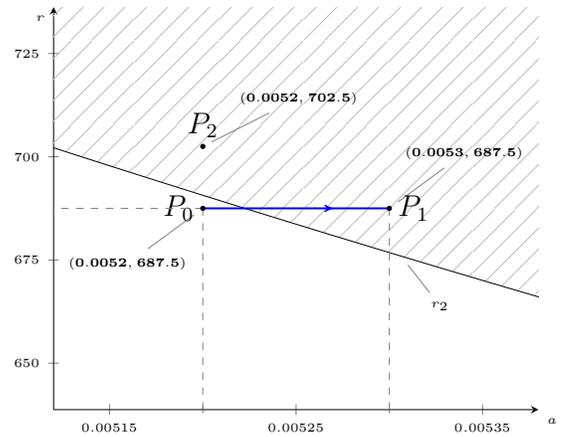}
	\caption{\label{fig:cont_proc:paths}
	$P_0$ : self-excited attractor, $P_1$ : hidden attractor, $P_2$ : no chaotic attracxtors.}
\end{figure}

One of the effective methods for numerical localization of hidden attractors
in multidimensional dynamical systems is based on a {\it homotopy}
and {\it numerical continuation}:
it is necessary to construct a sequence of
similar systems such that for the first (starting) system the initial
data for numerical computation of oscillating
solution (starting oscillation)
can be obtained analytically, e.g,
it is often possible to consider the starting
system with self-excited starting oscillation.
Then the transformation of this starting oscillation is tracked
numerically in passing  from one system to another.

Let us construct on the plane $(a,r)$  a line segment, intersecting
a boundary of the domain of stability of the equilibria $S_1$, $S_2$
(see Fig. \ref{fig:cont_proc:paths}).
Let us choose the point $P_1$ :
$r = 687.5$, $a = 0.0053$ as the finite point of the line segment.
To these parameters  correspond the following eigenvalues
of the equilibria of system~\eqref{sys:lorenz-general}:
\begin{eqnarray*}
S_0 : & 49.9619, \quad -1, \quad -54.9619 & \\
S_{1,2} : & -0.0968 \pm 10.4269 \mi, \quad -5.8063 &
\end{eqnarray*}
It means that the equilibria $S_{1,2}$ become stable focus-nodes.

Let us choose the point $$P_0:\ r= 687.5, a = 0.0052$$ as the initial point of the line segment.
This point corresponds to the parameters for which in system
\eqref{sys:lorenz-general} there exists a self-excited attractor, which
can be computed by the standard procedure.
Then for the considered line segment a sufficiently small partition step
is chosen and a chaotic attractor in the phase of system \eqref{sys:lorenz-general}
space at each iteration of the procedure is computed.
The last computed point at each step is used as the initial point
for the computation of the next step.


Our experiment has $8$ iterations and the partition
step equals $1.25 \cdot 10^{-5}$, respectively.
At each iteration for the current trajectory that describes the attractor
one computes the largest
Lyapunov exponent (${\rm LLE}$) \cite{BenettinGGS-1980}
and the Lyapunov dimension (${\rm LD}$) \cite{KaplanY-1979,BoichenkoLR-2005}~\footnote{
There are two widely used definitions of Lyapunov exponents:
the upper bounds of the exponential growth rate of the norms of linearized system solutions
and the upper bounds of the exponential growth rate of the singular values
of linearized system fundamental matrix.
While in typical case these two definitions gave the same values,
for a given system they may be different and
there are examples in which Benettin algorithm \cite{BenettinGGS-1980-Part2}
(see, e.g., its MATLAB implementation \cite{LET-1998})
fails to compute the correct values (see the discussion in \cite{KuznetsovAL-2014-arXiv-LE})}.

\begin{figure}[!hb]
 \centering
 \includegraphics[width=0.4\textwidth]{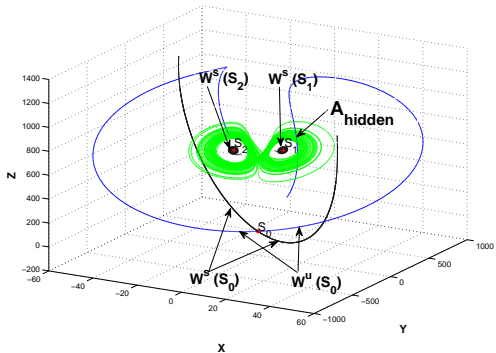}
 \caption{\label{fig:gen_lorenz:attr:hidden} Hidden attractor for system \eqref{sys:lorenz-general}.}
\end{figure}

Thus, for the selected path and selected partition
it is possible to visualize a hidden attractor of
system \eqref{sys:lorenz-general} (see Fig.~\ref{fig:gen_lorenz:attr:hidden}).

Remark that hidden attractor does not exist for all points of the shaded domain
in Fig.~\ref{fig:cont_proc:paths}.
E.g., there is no chaotic attractor for the point $P_2$ : $r = 702.5$, $a = 0.0052$.

Note also that in the work \cite{LeonovM-2015}
the upper estimation of the Lyapunov dimension (LD) of attractor
of system \eqref{sys:lorenz-general} is presented:
for $r = 687.5$, $a = 0.0053$, $\sigma = 4$ it was obtained analytically
the estimation ${\rm LD} < 2.8909$ that is in a good agreement with the numerical result
${\rm LD} = 2.1293$. 

\section{\label{sec:conclusions} Conclusions} 

In the present work by numerical methods
the scenario of  transition to chaos in physical model \eqref{sys:lorenz-general},
describing a flow of rotating fluid convection
inside the ellipsoid under horizontal heating, is demonstrated.
Similarly to scenario in the classical Lorenz system,
in system \eqref{sys:lorenz-general} a homoclinic trajectory
and self-excited chaotic attractor are constructed.
However, unlike the Lorenz system for system \eqref{sys:lorenz-general}
one is able to localize numerically a hidden attractor.

\begin{acknowledgments}
   This work was supported by Russian Scientific Foundation (project 14-21-00041)
   and Saint-Petersburg State University.
\end{acknowledgments}

%

\end{document}